\documentclass[12pt,preprint]{aastex}

\shorttitle{Herbig AeBe stars detections in X-Rays}
\shortauthors{Hamidouche et al.}

\begin{document}

\title{The X-Ray origin of Herbig AeBe Systems: New Insights}

\author{Murad Hamidouche\altaffilmark{1}, Shiya Wang, Leslie W. Looney} 
\affil{Astronomy Department, University of Illinois, Urbana, IL 61801}
\email{mhamidouche@sofia.usra.edu}

\altaffiltext{1}{New address: Stratospheric Observatory for 
Infrared Astronomy, NASA Ames Research Center, MS 211-3, Moffett Field, CA 94035}

\begin{abstract}
We present a statistical study of the X-Ray emission toward 22 Herbig
AeBe stars using the Chandra archive. We probe the origin of the X-Rays
toward Herbig stars: are they intrinsic? This question is addressed
by correlations between the physical stellar properties and the X-Ray
emission. There is a weak correlation between the continuum radio
emission at $\lambda$=3.6 cm and $L_X$, which suggests that the X-Ray
emission depends upon the source. On the other hand, no correlation
was found with the stellar rotational period, but that only excludes solar-like 
magnetic activity as the origin of the X-Rays. Most importantly,
the X-Ray luminosity of Herbig AeBe stars have a different distribution
than T Tauri stars, suggesting X-Ray emission from an unseen late type
star companion can be ruled out with an 80$\%$ confidence level. This implies that
the Herbig AeBe stars must have magnetic activity.
In addition, we report the observation of five sources for the first time, three detections.
\end{abstract}


\section{Introduction}

Herbig AeBe stars (HAEBE) are young intermediate mass stars, ranging
roughly from 2 to 20 M$_\odot$ of spectral type A, B and early F
\citep{her60}. They are considered the more massive counterparts of T
Tauri stars (TTS). The study of HAEBE disks has attracted particular
interest in investigating their formation and evolution into planetary
systems. Unfortunately, their pre-main-sequence (PMS) evolution is more
difficult to study with as much detail as TTS, as they less abundant and evolve faster;
the formation and evolution processes are presumably accelerated
and more embedded. The existence of circumstellar disks in HAEBE
systems were confirmed with interferometric millimeter observations
\citep[e.g.][]{mannings97}, and more recent results have resolved
these disks \citep[e.g.][]{mh06}. \citet{fuente03} have reported the
first evidence of a disk around the more massive Herbig Be stars. The
detection of these disks is relevant in probing the X-Ray origin in
HAEBE systems. In fact, numerous authors have already suggested star-disk
magnetic interactions may generate the X-Rays \citep{thm00}.

Although X-Ray detection toward HAEBE stars has become quite common
\citep[e.g.][]{hamaguchi05, stelzer06}, its origin is more difficult
to explain than the X-Ray detection of lower mass TTS. 
The later-type TTSs are also routinely observed in the
X-Ray. The process usually invoked to explain their X-Ray origin is
solar-type coronal loops in the 1-10 keV band \citep{F07}, while in some
active protostars, larger magnetic structures can possibly connect the
stellar photosphere and the circumstellar accreting disk \citep{thm00,
F07}. Using the Chandra Orion Ultradeep Project, COUP (Getman et
al. 2005a), and the XMM-Newton Extended Survey of the Taurus Molecular
Cloud, XEST \citep{G07b}, several studies have shown that TTS X-Rays are
mostly due to coronal emission and not to accretion emission, which only has
a minor effect in the soft $<$ 1 keV regime \citep[e.g.][]{P07,
G07b}. Nevertheless, accreting stars were found to have a lower X-Ray
activity than non-accreting sources. The accreting material cools down the
corona hot plasma. Therefore, the cool plasma generates very soft X-Rays
hardly detectable with Chandra and XMM-Newton \citep{P07}. 

Numerous surveys of X-Ray emission toward HAEBE stars have been done:
Stelzer et al. (2006) have reported a fraction of $\sim$76 $\%$ from
a sample of 17 sources using Chandra, including emission from known
companions; Hamaguchi et al. (2005) have detected 31$\%$ from a sample
of 35 sources using ASCA; Zinnecker \& Preibisch (1994) have detected
52$\%$ from a sample of 21 sources using ROSAT; and \citet{damiani94}
have detected 35$\%$ from a sample of 31 sources using Einstein. 

In this paper, we report on a sample of 22 sources using the Chandra archive,
17 sources from Stelzer et al. (2006) and 5 new sources.
With these data, we compare the X-Ray emission to the stellar properties in order to look for
possible correlations; hence, whether the X-Rays are intrinsic to the Herbig
stars or not.

\section{Data sample and Observations}\label{dat}

We have correlated the Chandra archive with the HAEBE stars catalogs of: Mora et al. (2001), Natta et
al. (2000), Th\'e et al. (1994), and Hillenbrand et al. (1992). We chose the
sources of spectral types earlier than F5. This provides a sample to probe and
compare the X-Rays of both the coolest Herbig Ae (late B, A and early F) stars
and the Herbig Be (early/mid B) stars (e.g. Natta et al. 2000). Data mining
the Chandra's public archive for observed sources allowed us to make a list of
22 HAEBE observations, whether they were serendipitous observations or not. In
fact, there are 13 sources that were not observed as the main target, but were
in the field of view during other observations. The list of sources with known
stellar parameters is given in Table 1. 

Fortuitously, our sample includes HAEBE sources of different spectral types,
which will provide a better analysis of the X-Ray properties of HAEBE
stars. Our sample includes stars
of masses ranging between $\sim$ 2-26 $M_\odot$ of different ages between 10$^6$-10$^7$ years \citep{vda98, vda00}.

Table 2 summarizes the Chandra observations. The observations are obtained
with the ACIS (Advanced CCD Imaging Spectrometer) detector within Chandra's
band 0.5 to 10 keV. 
Although many of our sources were reported in \cite{stelzer06}, the fact that we 
have 5 addition sources, lead us to re-analyze all of them in order to be consistent.
In addition, our detection criteria is more focused on the removal of companion emission
(see HD 150193 in \S \ref{detect}).
The Chandra data analysis software CIAO\footnote{The
  Chandra Interactive Analysis of Observations (CIAO) software package can be
  found at: http://asc.harvard.edu/ciao/} version 3.1 and the X-ray spectral
fitting package XSPEC version 11.3 are used for our data reduction and
analysis. We obtained the Event 2 level processed data from Chandra
Archives. We extract the background lightcurve and use ChIPS to exclude the
time periods of high background by removing spikes on the background
lightcurve. These processes give the good time interval. The event files with applied good time interval are then used to detect point sources and further extract spectra of these sources.

The WAVDETECT was used to detect point sources within a 50 $\times$ 50 pixel
area centered on the source IR/optical position. This algorithm uses Mexican Hat wavelet functions to correlate the images and identify source regions with large positive
correlation. We generate regions centered at the detected peak emission within
a 3$\sigma$ ellipse, including 90$\%$ of the point spread function (PSF) at 1.497 KeV. With detected point sources, we use Sherpa to plot the radial profile to look for any possible signatures of extra emission that may be due to companions.

\section{Results and discussions}

\subsection{X-Ray detections}\label{detect}

 Count rates are estimated within a circular or elliptical region,
depending on the source geometry, around the optical position of the source
after removing the background. The background is estimated in a
different region of the image map and normalized by the surface area of the
source region. Our criteria for the detection was based on the signal-to-noise
ratio (S/N), S/N higher than $\approx$3. The count rate was estimated in the
``broad'' band (0.5-8 keV), including both soft and hard X-Rays. The sources
with low S/N were classified as non-detections.

The number of detected sources is 14, out of 22 sources; 8 of the detected
sources were serendipitous observations of Chandra, 3 of which had not yet been reported. 
Table 3 summarizes all the
sources and their corresponding X-Ray luminosity. The upper limits
were estimated at a 90$\%$ confidence level using the Bayesian method
\citep{kraft91}. The deduced percentage of HAEBE stars detected $\simeq$ 64
$\%$ (see Table 3), compared to 76$\%$ from \cite{stelzer06}. 
Out of the 5 Herbig Be stars, earlier than B6, 3 were detected (60$\%$), and out of the 17 Herbig Ae stars, 11 were detected (65$\%$). 

For the 8 brightest detected sources, the flux
density was derived by fitting thermal plasma models based on
MEKAL\footnote{Mekal = Mewe-Kaastra-Liedahl thermal plasma (1995);
  http://cxc.harvard.edu/ciao/ahelp/xs.html} emissivities \citep{km00} (Table
4). The flux density, gas column density N$_{H_2}$, and the gas temperature kT
are then derived from this spectral analysis. 
Extrapolating linearly the flux density values of the
8 brightest sources against the count rates allows an estimation of the flux
density for the other 6 sources, which have lower count rates. N$_{H_2}$ from
the spectra fitting could be used to correct our flux from the absorption of the gas through the line-of-sight. However, we chose not to correct our flux since these
N$_{H_2}$ values are not well determined \citep[e.g.][]{stassun04, feigelson02}. The X-Ray luminosity ($L_X$) were then computed for each
source by multiplying the total flux density in the entire band 0.5-8 keV by
4$\pi d^2$, where $d$ is the distance of the source (Table 1). In addition to
the count rate uncertainties, we consider distance uncertainties to estimate
$L_X$ uncertainties. If distance uncertainties are not presented in the
literature, we use a 20$\%$ uncertainty.

We can compare our results to those of \citet{stelzer06}.
The deduced Log $L_X$ are consistent within the uncertainties.
However, AK Sco shows a very low count-rate below our detection
criteria; we do not consider it as detected. We have also detected X-Ray
emission toward HD 150193, but it is offset by $\simeq$ 2.5\arcsec~ from its
optical position (cf. Stelzer et al. 2006). These sources were both considered
detected by Stelzer et al. (2006). 

In addition, this is the first study where
Chandra observations toward the additional five sources (underlined in Table
3) are reported (Figure 1). 
From these five sources, we detected V361 Ori, AB Aur, and V372 Ori, and did not detect LP Ori and MR Ori. Using ASCA, Hamaguchi et al. (2005) also did
not detect LP Ori or MR Ori. In the ROSAT survey, \citet{zinnecker94} detected AB Aur with a slightly higher
Log $L_X$ = 29.5 ergs/sec, while it was not detected by Hamaguchi et
al. (2005) and \citet{damiani94} using Einstein. V372 Ori was detected by
\citet{gagne95} in a ROSAT survey, Log $L_X$ = 30.3 ergs/sec. These
results are consistent with our detections; however, our $L_X$
are slightly lower since we did not correct for absorption. This is the first
detection of V361 Ori.

\subsection{X-Ray relations to stellar properties}\label{stars}

If the X-ray emission is intrinsic to the stellar systems, there may be some correlation
between the star and the X-Ray emission.
We have performed Kendall's $\tau$-tests, including the
upper limits data (non-detections), as implemented in the ASURV package \citep{isobe86}. We compare first the stellar bolometric luminosity
with the X-Ray luminosity $L_X$. We find a mean ratio Log
($L_X/L_{bol}$) = -5.62$\pm$1.18 for the detected sources (Table 3). This ratio is consistent with the recent
values found toward HAEBE stars (e.g. Skinner et
al. 2004). Figure 2 (\textit{top}) shows $L_X$, for detected sources and
upper limits for the undetected ones, versus $L_{bol}$. Most points are between the two lines corresponding to
the TTS ratio -3.75 \citep{skinner04} and main sequence OB stars ratio -7.0
\citep{berg97}. Using the Kendall's $\tau$-test, we found a probability of $P$ =
0.42 that a correlation is not present. Interestingly, the test for the surface area ($4 \pi r_*^2$ =
$L_{bol}$/$\sigma T_{eff}^4$) and $L_X$ provides a similar result $P$ = 0.39
(Figure 2, \textit{bottom}).

We probe the relations of $L_X$ with the stellar rotation period,
$P_{rot}$=$2\pi r_*/v_{rot} sin i$, and the wind velocity $v_{wind}$, for
the sources with known values (Table 1). We did not find a correlation
between the luminosity ratio $L_X/L_{bol}$ and the stellar rotational
period $P_{rot}$, Kendall $\tau$-test probability of no correlation $P$
= 0.44 (Figure 3).
For comparison, late type main-sequence stars are known to have a clear
  correlation between the luminosity ratio and the stellar rotation
  period \citep{pallavicini81, P07}.  On the other hand, they  did not use their
  non-detection limits, so  we repeated the Kendall $\tau$-test for
  only our detected sources and still did not find a strong correlation, probability
  of no correlation $P$ = 0.65.
Therefore, solar-like magnetic dynamo mechanism can be excluded as the
origin of HAEBE X-Ray activity.
 We used the relation of mass-loss rate and bolometric luminosity
for Herbig stars given by \citep{skinner93}: Log $\dot{M}$ = -9.1 +
0.6 Log ($L_{bol}/L_{\odot}$) M$_\odot$/year to deduce the wind kinetic
luminosity $L_{kin}$=1/2 $\dot{M}$ v$_{wind}^2$. Figure 4 shows that
$L_{kin}$ is below the dashed line, corresponding
  to $L_X$ = $L_{kin}$, by about two order of magnitude. In addition,
  Figure 5 shows that most points are above the maximum temperature,
  dashed line, that can be generated if all the kinetic energy is
converted into thermal bremsstrahlung energy. Although, only three
sources have known $v_{wind}$ and do not satisfy this later condition,
we can suggest
  that the wind-shock model does not appear as the origin of the X-Ray
  emission for these
sources. HD 104237 has a wind velocity that may generate part of
the X-Rays. Its corresponding point is below the dashed line (Figure
5). \citet{skinner04}
  have suggested a possible existence in this source of a thin convective
  zone of $\approx$ 0.9$\%$ stellar radius and a magnetic activity,
  but it may not be strong enough to produce the detected $L_X$. We
  can naively suggest that a fraction of the wind kinetic energy can be
  a complementary process to the stellar coronal magnetic activity to
  produce the observed X-Rays.

Figure 6 displays $L_X$ versus the radio continuum luminosity at $\lambda$ =
3.6 cm, $L_{3.6 cm}$. There is a correlation between the two
variables. Kendall $\tau$-test's probability of no correlation of only $P$ =
0.025, which becomes $P$=0.01 when only the detected sources in both
X-Ray and radio are considered in the test. We
deduce an almost linear relation between $L_X$ and the stellar radio emission $L_X
\propto 10^{11-12} L_{3.6 cm}$[Hz]. Both X-Ray and radio emissions can be related to the stellar
magnetic activity at different levels, if assuming that they come from the same
star. Similarly, \citet{gudel02} reported in his analysis toward active stars
with a hot plasma emitting both thermal X-Rays and non-thermal radio
radiation. However, we note that our analysis is for Herbig Ae
stars only, for which we know
$L_{3.6 cm}$, and they are known to possibly have a thin convective zone that
may generate the X-Rays \citep[e.g.][]{skinner04}.

\section{X-Ray Emission from Companions?}\label{tts}

The X-Ray emission from intermediate mass HAEBE stars (and AB stars) has
been a standing puzzle as they are not known to have convective outer layers
that generate the magnetic dynamo as in the lower mass TTS \citep[e.g.][]{fm99}. The most common explanation of
the X-Ray origin is from an unresolved lower mass TTS companion
(e.g. Zinnecker \& Preibisch 1994; Feigelson et al. 2003). However, Chandra
can not resolve companions closer than $\simeq$ 1\arcsec, or $\sim$ 100-1000
AU for our sources. This is larger than the typical binary separations, which
can be $\sim$ 0.1\arcsec~ \citep[e.g.][]{baines06, tokovinin06}.

\subsection{Comparison to the Orion Nebula sources}\label{ONC}

We combine our detected HAEBE stars data with TTS and HAEBE
stars observed in the ONC. The Orion observations used here are from the COUP project. COUP has detected more than
1600 X-Ray sources of different spectral types and ages $\sim$ 10$^{4.2}$ -
$10^{7.6}$ years, a similar age range to our sources. We select COUP stars that have
known spectral types. Since COUP observations are much
more sensitive than Chandra observations in our sample, we also truncated the
COUP sample at Log $L_X >$ 28.59 corresponding to the lowest $L_X$ of our
sample. This prevents a comparison of inhomogeneous observations in terms of
sensitivity limit. Making a list with the ONC sources and completing it with our HAEBE
stars give us a unique opportunity to directly compare the X-Ray luminosity
distribution of different spectral type stars. To make
our statistical comparison consistent, we use their uncorrected $L_X$. This
should not affect the ensemble statistics. 

The sources were sampled into 3 groups of stellar objects: 1) Group I
earlier than B3, 2) Group II intermediate mass stars, HAEBE, of
spectral type B3-F5, and 3) Group III TTS, spectral type later
than F5. We chose to use the spectral type
to make different groups instead of the mass since the spectral type is often
better known. However, we checked the known masses of Group III sources
(or TTS) and found that their masses $\lesssim$ 3$M_\odot$ are consistent with
TTS.

Figure 7 shows the X-Ray luminosity variation with the spectral type. In Group I, the X-Ray luminosity is very scattered, $\sigma$$_{Log L_X}$ =
1.26, around the mean value \textit{Log} $L_X$ = 30.74. Group II sources are
less scattered, $\sigma$$_{Log L_X}$ = 0.82, and have a slightly lower mean
value \textit{Log} $L_X$ = 30.1. We also note in Group II that the $L_X$ range of
our sources is similar to the range of intermediate-mass stars from the COUP observations. There is no apparent dependence of $L_X$
with spectral type in both Group I and II. On the other hand, Group III shows
a slight dependence of the luminosity with spectral type. It decreases
with the spectral type. The luminosity mean value
is Log $L_X$ = 29.76, smaller than in Group I and slightly smaller than in
Group II. Figure 8 shows the $L_X$ cumulative distribution
 function of TTS (Group III), Herbig Ae, Herbig Be,
 and HAEBE (Group II) samples. The $L_X$ distribution for Group III follows
 a nearly uniform distribution. Herbig Ae's curve mostly resembles Group
 III's; but Group III's curve has a more extended tail toward lower $L_X$. Clearly the Herbig
 Be distribution is very different from the others. This is expected due to the very
 large scattered $L_X$ (see Figure 7).

TTS' X-Ray luminosity shows some dependence on spectral type while
the other groups do not. The detected X-Rays of the HAEBE stars may have a
different origin than TTS companion, or the HAEBE stars do not emit in
the X-Rays but have similar X-Ray emitting companions, hence an almost
constant $L_X$. Nevertheless, some single HAEBEs have already shown intrinsic
X-Ray emission \citep[e.g.][]{swartz05}. Thus, the
detected $L_X$ may be a sum of the emission from the Herbig source and a
possible companion's emission. 

\subsection{Statistical comparison}\label{stat}

To quantify our finding, we use a Kolmogorov-Smirnov (K-S) 
\citep[e.g.][]{press93} and the Wilcoxon
rank-sumtest (WRS) tests \citep{lehmann75} to test the $L_X$
distributions. To check the
robustness of this statistical comparison, we
split randomly the largest sample, Group III, into 2
sub-Groups (test1 in Table 5). We find that the two sub-Groups derive from the
same distribution, with a confidence level higher than 99.99$\%$ (K-S) and
95$\%$ (WRS). This shows that our strategy is effective.

The results are presented in Table 5. The two tests (K-S and WRS)
provide a consistent variation of the probability, WRS probabilities are
lower, which may be due to the difference of the median values of $L_X$ in each
Group. 

\subsection{Discussion}\label{disc}

We can rule out the hypothesis that X-Rays detected toward Herbig systems
(Group II) are from TTS companions with an 80$\%$ confidence level. Furthermore, we can
not reject the hypothesis that Herbig Ae stars and TTS X-Rays derive from a same distribution. The probability, 12$\%$, that Herbig Ae stars' X-Rays having a different parent distribution than TTS is much lower than the entire Herbig stars ensemble. This may be due to the fact that the process generating
the X-Rays in the Herbig Ae stars is similar to the TTS's. In the same
way, it was already proposed that late Herbig Ae stars may have an outer
convective zone that supports the magnetic activity,
similar but quite thinner than TTS' (e.g. Skinner et al. 2004). Herbig stars
and OB stars (Group I) have a same parent distribution at a 55$\%$
confidence level. This may be due to a similar origin of the X-Rays, magnetic
activity caused by a fossil magnetic field from the parent molecular
cloud. This may be the case for at least the more massive Herbig Be stars,
which have a relatively different $L_X$ distribution than the Herbig Ae stars
(see \S \ref{stat}), and evolve faster than the Herbig Ae stars
\citep{natta00}. It is important to note that the uncertainties of the source spectral type will
affect the group selection, particularly at the 
edges around F5 or B2.  When placing sources in the group, we compared their
masses (e.g. Getman et al. 2005b) before placing them in the respective group.
However, this is not a large effect on our statistical tests as the key parameter is the overall 
distribution of the $L_X$ over the group, not the exact placement in the
group. \citet{stelzer06} reported that the X-Ray emission from HAEBE stars
was qualitatively more similar to TTS in ONC than main sequence OB stars.
However, this was based on a plot of $L_X$ versus $kT$ and was not quantified. 

\section{Summary}

 The relevance of our X-Ray study of HAEBE stars is due to our
large sample with a somewhat arbitrary source selection; more than half of the sources were serendipitous observations. In addition, this is the first time where such a large sample of HAEBE stars are investigated in X-Rays with the high resolution of Chandra. The main results of this study are:

1) Out of 22 HAEBE sources, 14 have been detected in X-Rays, about
   64$\%$. The Herbig Ae stars have a higher rate of
   detection compared to the Herbig Be stars. The luminosity
   ranges between Log $L_X$ = 30-31 (ergs/s). This is higher than TTS but
   overlaps the TTS $L_X$ range. We report the first detection of V361 Ori and the first detections of AB Aur and V372 Ori with Chandra. 

2) Although, the wind kinetic energy is strong enough to produce the detected $L_X$, the estimated temperatures are relatively high to be generated by such low-velocities. This shows that the wind shock model does not appear to generate the observed X-Rays, but one needs to be careful as this statement is based on only a few sources. More investigation including more sources with known $v_{wind}$ need to be done to confirm this. Nevertheless, HD 104237 has a relatively high wind velocity. Its X-Ray emission can be partially or fully due to the kinetic $L_{kin}$.   

3) Comparing the X-Ray emission to the stellar parameters: i) A luminosity
   ratio Log $L_X/L_{bol}$ = -5.62$\pm$1.18 is lower than the typical
   TTS's ratio; ii) There is no correlation with the rotational period
   $P_{rot}$, which
   excludes the possibility of a solar-like dynamo effects to produce the
   X-Rays for the Herbig stars for which we know \textit{v sin i}; iii) We deduced a nearly linear correlation
   between the continuum radio emission at $\lambda$=3.6 cm and $L_X$ toward Herbig Ae
   sources, with known $L_{3.6 cm}$. Thus, suggesting that the emission does not depend on a companion. 

4) The results of \S \ref{tts} show that HAEBE stars' X-Ray
   emission being from an unresolved TTS companion can be ruled out at an 80$\%$
   confidence level using the K-S test on the $L_X$ distribution. In addition,
   the results show that the X-Ray emission of HAEBE
   stars is
   different than OB stars with only a 45$\%$ confidence level.

Overall, we suggest that the X-Rays are intrinsic to the HAEBE stars. In that case, they must have stellar
magnetic activity. This is likely due to the remnant magnetic field after the
collapse \citep{tassis04, mont05}. Indeed
\citet{wade07} using spectropolarimeter observations reported the measurement of a magnetic
field toward HAEBE stars. The existence of circumstellar disks toward HAEBE
systems have been confirmed observationally in the last few years
\citep[e.g.][]{natta00}. Star-disk magnetic interaction can be an appropriate
explanation of the X-Ray origin (e.g. Montmerle et al. 2000).

\acknowledgments

We are grateful to the anonymous referee for the several valuable
suggestions to improve the paper. We thank Thierry Montmerle for his helpful comments on the paper. We acknowledge support from the Laboratory for Astronomical Imaging at the University of Illinois and NSF AST 0228953. We would like to thank Alastair Sanderson and Rosa Williams for their interesting comments. We would like to thank Alessandro Gardini for the several discussions on physical processes in the X-Ray.    


\begin{thebibliography}{53}
\expandafter\ifx\csname natexlab\endcsname\relax\def\natexlab#1{#1}\fi
\expandafter\ifx\csname url\endcsname\relax
  \def\url#1{\texttt{#1}}\fi
\expandafter\ifx\csname urlprefix\endcsname\relax\def\urlprefix{URL }\fi
\providecommand{\eprint}[2][]{\url{#2}}

\bibitem[{{Arnaud}(1996)}]{arnaud96}
{Arnaud}, K.~A. 1996, in ASP Conf. Ser. 101: Astronomical Data Analysis
  Software and Systems V, edited by G.~H. {Jacoby}, \& J.~{Barnes}, 17

\bibitem[{{Baines} et~al.(2006){Baines}, {Oudmaijer}, {Porter}, \&
  {Pozzo}}]{baines06}
{Baines}, D., {Oudmaijer}, R.~D., {Porter}, J.~M., \& {Pozzo}, M. 2006, \mnras,
  367, 737. \eprint{astro-ph/0512534}

\bibitem[{{Berghoefer} et~al.(1997){Berghoefer}, {Schmitt}, {Danner}, \&
  {Cassinelli}}]{berg97}
{Berghoefer}, T.~W., {Schmitt}, J.~H.~M.~M., {Danner}, R., \& {Cassinelli},
  J.~P. 1997, \aap, 322, 167




\bibitem[{{Damiani} et~al.(1994){Damiani}, {Micela}, {Sciortino}, \&
  {Harnden}}]{damiani94}
{Damiani}, F., {Micela}, G., {Sciortino}, S., \& {Harnden}, F.~R. 1994, \apj,
  436, 807


\bibitem[{{Feigelson} et~al.(2007){Feigelson}, {Townsley}, {G{\"u}del}, \&
  {Stassun}}]{F07}
{Feigelson}, E., {Townsley}, L., {G{\"u}del}, M., \& {Stassun}, K. 2007, in
  Protostars and Planets V, edited by B.~{Reipurth}, D.~{Jewitt}, \& K.~{Keil},
  313


\bibitem[{{Feigelson} et~al.(2003){Feigelson}, {Lawson}, \& {Garmire}}]{F03}
{Feigelson}, E.~D., {Lawson}, W.~A., \& {Garmire}, G.~P. 2003, \apj, 599, 1207

\bibitem[{{Feigelson} et~al.(2002){Feigelson}, {Broos}, {Gaffney}, {Garmire},
  {Hillenbrand}, {Pravdo}, {Townsley}, \& {Tsuboi}}]{feigelson02}
{Feigelson}, E.~D., {Broos}, P., {Gaffney}, J.~A., III, {Garmire}, G.,
  {Hillenbrand}, L.~A., {Pravdo}, S.~H., {Townsley}, L., \& {Tsuboi}, Y. 2002,
  \apj, 574, 258. \eprint{astro-ph/0203316}

\bibitem[{{Feigelson} \& {Montmerle}(1999)}]{fm99}
{Feigelson}, E.~D., \& {Montmerle}, T. 1999, \araa, 37, 363

\bibitem[{{Forbrich} et~al.(2006){Forbrich}, {Preibisch}, \&
  {Menten}}]{forbrich06}
{Forbrich}, J., {Preibisch}, T., \& {Menten}, K.~M. 2006, \aap, 446, 155

\bibitem[{{Fuente} et~al.(2002){Fuente}, {Mart{\i}n-Pintado}, {Bachiller},
  {Rodr{\i}guez-Franco}, \& {Palla}}]{fuente02}
{Fuente}, A., {Mart{\i}n-Pintado}, J., {Bachiller}, R., {Rodr{\i}guez-Franco},
  A., \& {Palla}, F. 2002, \aap, 387, 977

\bibitem[{{Fuente} et~al.(2003){Fuente}, {Rodr{\'{\i}}guez-Franco}, {Testi},
  {Natta}, {Bachiller}, \& {Neri}}]{fuente03}
{Fuente}, A., {Rodr{\'{\i}}guez-Franco}, A., {Testi}, L., {Natta}, A.,
  {Bachiller}, R., \& {Neri}, R. 2003, \apjl, 598, L39

\bibitem[{{Gagne} et~al.(1995){Gagne}, {Caillault}, \& {Stauffer}}]{gagne95}
{Gagne}, M., {Caillault}, J.-P., \& {Stauffer}, J.~R. 1995, \apj, 445, 280

\bibitem[{{Getman} et~al.(2005{\natexlab{a}}){Getman}, {Feigelson}, {Grosso},
  {McCaughrean}, {Micela}, {Broos}, {Garmire}, \& {Townsley}}]{getman05a}
{Getman}, K.~V., {Feigelson}, E.~D., {Grosso}, N., {McCaughrean}, M.~J.,
  {Micela}, G., {Broos}, P., {Garmire}, G., \& {Townsley}, L.
  2005{\natexlab{a}}, \apjs

\bibitem[{{Getman} et~al.(2005{\natexlab{b}}){Getman}, {Flaccomio}, {Broos},
  {Grosso}, {Tsujimoto}, {Townsley}, {Garmire}, {Kastner}, {Li}, {Harnden},
  {Wolk}, {Murray}, {Lada}, {Muench}, {McCaughrean}, {Meeus}, {Damiani},
  {Micela}, {Sciortino}, {Bally}, {Hillenbrand}, {Herbst}, {Preibisch}, \&
  {Feigelson}}]{getman05b}
{Getman}, K.~V., {Flaccomio}, E., {Broos}, P.~S., {Grosso}, N., {Tsujimoto},
  M., {Townsley}, L., {Garmire}, G.~P., {Kastner}, J., {Li}, J., {Harnden},
  F.~R., {Wolk}, S., {Murray}, S.~S., {Lada}, C.~J., {Muench}, A.~A.,
  {McCaughrean}, M.~J., {Meeus}, G., {Damiani}, F., {Micela}, G., {Sciortino},
  S., {Bally}, J., {Hillenbrand}, L.~A., {Herbst}, W., {Preibisch}, T., \&
  {Feigelson}, E.~D. 2005{\natexlab{b}}, \apjs

\bibitem[{{G{\"u}del}(2002)}]{gudel02}
{G{\"u}del}, M. 2002, \araa, 40, 217. \eprint{astro-ph/0206436}

\bibitem[{{G{\"u}del}(2007)}]{G07b}
{G{\"u}del}, M. 2007, Memorie della Societa Astronomica Italiana, 78, 422

\bibitem[{{Hamaguchi} et~al.(2005){Hamaguchi}, {Yamauchi}, \&
  {Koyama}}]{hamaguchi05}
{Hamaguchi}, K., {Yamauchi}, S., \& {Koyama}, K. 2005, \apj, 618, 360

\bibitem[{{Hamidouche} et~al.(2006){Hamidouche}, {Looney}, \& {Mundy}}]{mh06}
{Hamidouche}, M., {Looney}, L.~W., \& {Mundy}, L.~G. 2006, \apj, 651, 321.
  \eprint{astro-ph/0607232}

\bibitem[{{Herbig}(1960)}]{her60}
{Herbig}, G.~H. 1960, \apjs, 4, 337

\bibitem[{{Hillenbrand} et~al.(1992){Hillenbrand}, {Strom}, {Vrba}, \&
  {Keene}}]{hillenbrand92}
{Hillenbrand}, L.~A., {Strom}, S.~E., {Vrba}, F.~J., \& {Keene}, J. 1992, \apj,
  397, 613

\bibitem[{{Isobe} et~al.(1986){Isobe}, {Feigelson}, \& {Nelson}}]{isobe86}
{Isobe}, T., {Feigelson}, E.~D., {Nelson}, P.~I. 1986, \apj, 306, 490

\bibitem[{{Kaastra} \& {Mewe}(2000)}]{km00}
{Kaastra}, J.~S., \& {Mewe}, R. 2000, in Atomic Data Needs for X-ray Astronomy,
  p. 161, edited by M.~A. {Bautista}, T.~R. {Kallman}, \& A.~K. {Pradhan}, 161

\bibitem[{{Kraft} et~al.(1991){Kraft}, {Burrows}, \& {Nousek}}]{kraft91}
{Kraft}, R.~P., {Burrows}, D.~N., \& {Nousek}, J.~A. 1991, \apj, 374, 344

\bibitem[{{Lehmann}(1975)}]{lehmann75}
{Lehmann}, E.~L. 1975, in Nonparametrics: Statistical methods based on ranks,
  edited by N.~Y. McGraw-Hill

\bibitem[{{Malfait} et~al.(1998){Malfait}, {Bogaert}, \&
  {Waelkens}}]{malfait98}
{Malfait}, K., {Bogaert}, E., \& {Waelkens}, C. 1998, \aap, 331, 211

\bibitem[{{Mannings} \& {Sargent}(1997)}]{mannings97}
{Mannings}, V., \& {Sargent}, A.~I. 1997, \apj, 490, 792

\bibitem[{{Montmerle} et~al.(2000){Montmerle}, {Grosso}, {Tsuboi}, \&
  {Koyama}}]{thm00}
{Montmerle}, T., {Grosso}, N., {Tsuboi}, Y., \& {Koyama}, K. 2000, \apj, 532,
  1097

\bibitem[{{Montmerle} et~al.(2005){Montmerle}, {Wade}, {Landstreet},
  {M{\`e}nard}, {Grosso}, \& {Feigelson}}]{mont05}
{Montmerle}, T., {Wade}, G., {Landstreet}, J., {M{\`e}nard}, F., {Grosso}, N.,
  \& {Feigelson}, E.~D. 2005, in Protostars and Planets V, Proceedings of the
  Conference held October 24-28, 2005, in Hilton Waikoloa Village, Hawai'i. LPI
  Contribution No. 1286., p.8112, 8112

\bibitem[{{Mora} et~al.(2001){Mora}, {Mer{\'{\i}}n}, {Solano}, {Montesinos},
  {de Winter}, {Eiroa}, {Ferlet}, {Grady}, {Davies}, {Miranda}, {Oudmaijer},
  {Palacios}, {Quirrenbach}, {Harris}, {Rauer}, {Cameron}, {Deeg},
  {Garz{\'o}n}, {Penny}, {Schneider}, {Tsapras}, \& {Wesselius}}]{mora01}
{Mora}, A., {Mer{\'{\i}}n}, B., {Solano}, E., {Montesinos}, B., {de Winter},
  D., {Eiroa}, C., {Ferlet}, R., {Grady}, C.~A., {Davies}, J.~K., {Miranda},
  L.~F., {Oudmaijer}, R.~D., {Palacios}, J., {Quirrenbach}, A., {Harris},
  A.~W., {Rauer}, H., {Cameron}, A., {Deeg}, H.~J., {Garz{\'o}n}, F., {Penny},
  A., {Schneider}, J., {Tsapras}, Y., \& {Wesselius}, P.~R. 2001, \aap, 378,
  116

\bibitem[{{Natta} et~al.(2000){Natta}, {Grinin}, \& {Mannings}}]{natta00}
{Natta}, A., {Grinin}, V., \& {Mannings}, V. 2000, Protostars and Planets IV,
  559

\bibitem[{{Natta} et~al.(2004){Natta}, {Testi}, {Neri}, {Shepherd}, \&
  {Wilner}}]{natta04}
{Natta}, A., {Testi}, L., {Neri}, R., {Shepherd}, D.~S., \& {Wilner}, D.~J.
  2004, \aap, 416, 179. \eprint{astro-ph/0311624}


\bibitem[{{Pallavicini} et~al.(1981){Pallavicini}, {Golub}, {Rosner}, {Vaiana},
  {Ayres}, \& {Linsky}}]{pallavicini81}
{Pallavicini}, R., {Golub}, L., {Rosner}, R., {Vaiana}, G.~S., {Ayres}, T., \&
  {Linsky}, J.~L. 1981, \apj, 248, 279

\bibitem[{{Preibisch}(2007)}]{P07} 
{Preibisch}, T. 2007, Memorie della Societa Astronomica Italiana, 78, 332


\bibitem[{{Press} et~al.(1993){Press}, {Teukolsky}, {Vetterling}, {Flannery},
  {Lloyd}, \& {Rees}}]{press93}
{Press}, W.~H., {Teukolsky}, S.~A., {Vetterling}, W.~T., {Flannery}, B.~P.,
  {Lloyd}, C., \& {Rees}, P. 1993, Numerical recipes in Fortran: the art of
  scientific computing / Cambridge U Press

\bibitem[{{Rodr{\'{\i}}guez} et~al.(2007){Rodr{\'{\i}}guez}, {Zapata}, \&
  {Ho}}]{rod07}
{Rodr{\'{\i}}guez}, L.~F., {Zapata}, L., \& {Ho}, P.~T.~P. 2007, Revista
  Mexicana de Astronomia y Astrofisica, 43, 149. \eprint{astro-ph/0611754}

\bibitem[{{Skinner} et~al.(1993){Skinner}, {Brown}, \& {Stewart}}]{skinner93}
{Skinner}, S.~L., {Brown}, A., \& {Stewart}, R.~T. 1993, \apjs, 87, 217

\bibitem[{{Skinner} et~al.(2004){Skinner}, {G{\"u}del}, {Audard}, \&
  {Smith}}]{skinner04}
{Skinner}, S.~L., {G{\"u}del}, M., {Audard}, M., \& {Smith}, K. 2004, \apj,
  614, 221

\bibitem[{{Stassun} et~al.(2004){Stassun}, {Ardila}, {Barsony}, {Basri}, \&
  {Mathieu}}]{stassun04}
{Stassun}, K.~G., {Ardila}, D.~R., {Barsony}, M., {Basri}, G., \& {Mathieu},
  R.~D. 2004, \aj, 127, 3537. \eprint{astro-ph/0403159}

\bibitem[{{Stauffer} et~al.(1994){Stauffer}, {Caillault}, {Gagne}, {Prosser},
  \& {Hartmann}}]{stauffer94}
{Stauffer}, J.~R., {Caillault}, J.-P., {Gagne}, M., {Prosser}, C.~F., \&
  {Hartmann}, L.~W. 1994, \apjs, 91, 625

\bibitem[{{Stelzer} et~al.(2006){Stelzer}, {Micela}, {Hamaguchi}, \&
  {Schmitt}}]{stelzer06}
{Stelzer}, B., {Micela}, G., {Hamaguchi}, K., \& {Schmitt}, J.~H.~M.~M. 2006,
  \aap, 457, 223. \eprint{astro-ph/0605590}

\bibitem[{{Swartz} et~al.(2005){Swartz}, {Drake}, {Elsner}, {Ghosh}, {Grady},
  {Wassell}, {Woodgate}, \& {Kimble}}]{swartz05}
{Swartz}, D.~A., {Drake}, J.~J., {Elsner}, R.~F., {Ghosh}, K.~K., {Grady},
  C.~A., {Wassell}, E., {Woodgate}, B.~E., \& {Kimble}, R.~A. 2005, \apj, 628,
  811

\bibitem[{{Tassis} \& {Mouschovias}(2004)}]{tassis04}
{Tassis}, K., \& {Mouschovias}, T.~C. 2004, \apj, 616, 283


\bibitem[{{Th\'e} et~al.(1994){Th\'e}, {de Winter}, \& {Perez}}]{the94}
{Th\'e}, P.~S., {de Winter}, D., \& {Perez}, M.~R. 1994, \aaps, 104, 315

\bibitem[{{Tokovinin} et~al.(2006){Tokovinin}, {Thomas}, {Sterzik}, \&
  {Udry}}]{tokovinin06}
{Tokovinin}, A., {Thomas}, S., {Sterzik}, M., \& {Udry}, S. 2006, \aap, 450,
  681. \eprint{astro-ph/0601518}

\bibitem[{{van den Ancker} et~al.(2000){van den Ancker}, {Bouwman},
  {Wesselius}, {Waters}, {Dougherty}, \& {van Dishoeck}}]{vda00}
{van den Ancker}, M.~E., {Bouwman}, J., {Wesselius}, P.~R., {Waters},
  L.~B.~F.~M., {Dougherty}, S.~M., \& {van Dishoeck}, E.~F. 2000, \aap, 357,
  325

\bibitem[{{van den Ancker} et~al.(1998){van den Ancker}, {de Winter}, \& {Tjin
  A Djie}}]{vda98}
{van den Ancker}, M.~E., {de Winter}, D., \& {Tjin A Djie}, H.~R.~E. 1998,
  \aap, 330, 145

\bibitem[{{Wade} et~al.(2007){Wade}, {Bagnulo}, {Drouin}, {Landstreet}, \&
  {Monin}}]{wade07}
{Wade}, G.~A., {Bagnulo}, S., {Drouin}, D., {Landstreet}, J.~D., \& {Monin}, D.
  2007, \mnras, 376, 1145

\bibitem[{{Zinnecker} \& {Preibisch}(1994)}]{zinnecker94}
{Zinnecker}, H., \& {Preibisch}, T. 1994, \aap, 292, 152

\end{thebibliography}

\clearpage
\begin{figure}
  \begin{center}
    \mbox{
      {\includegraphics[angle=0,scale=0.43]{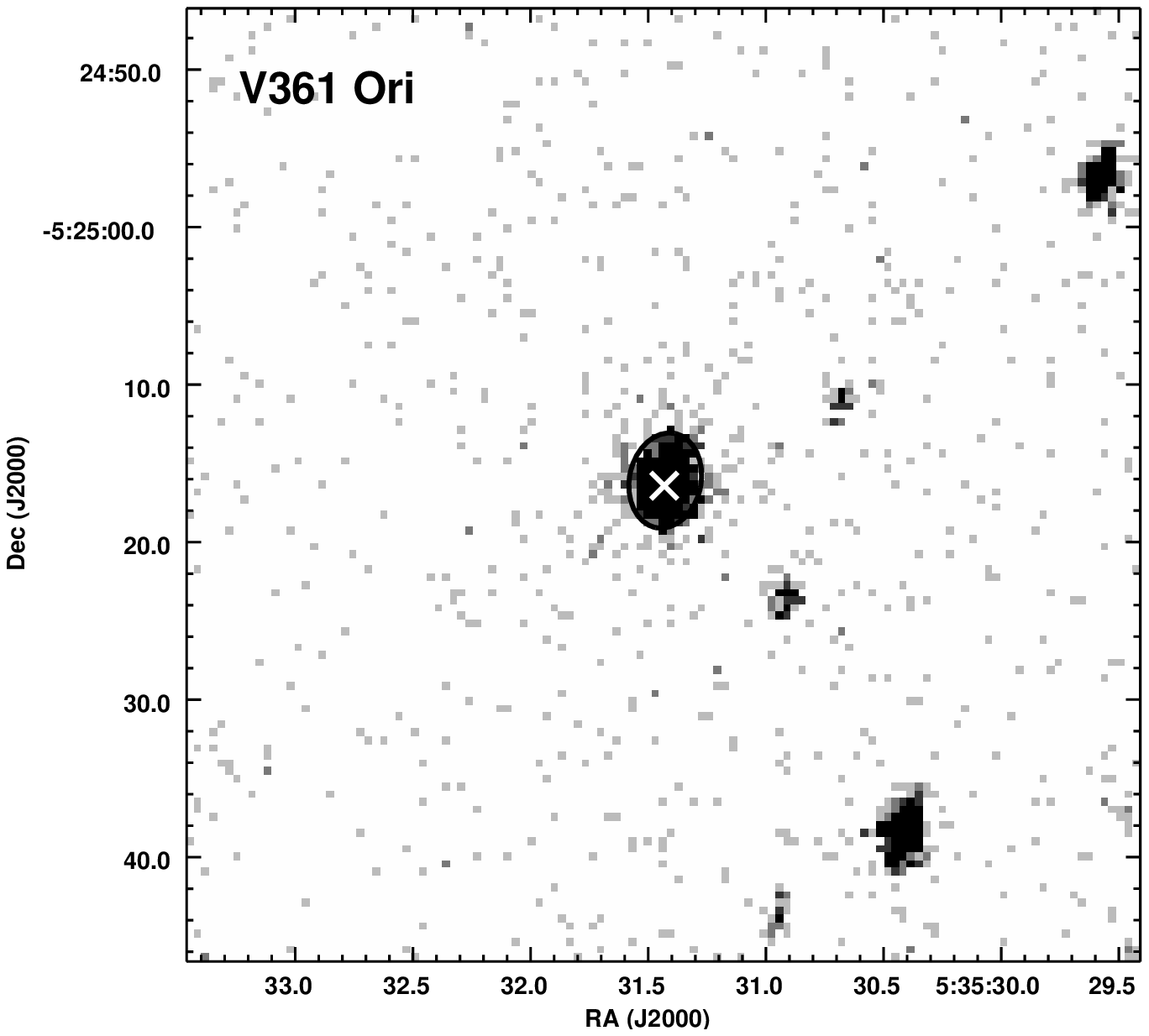}}\quad
      {\includegraphics[angle=0,scale=0.43]{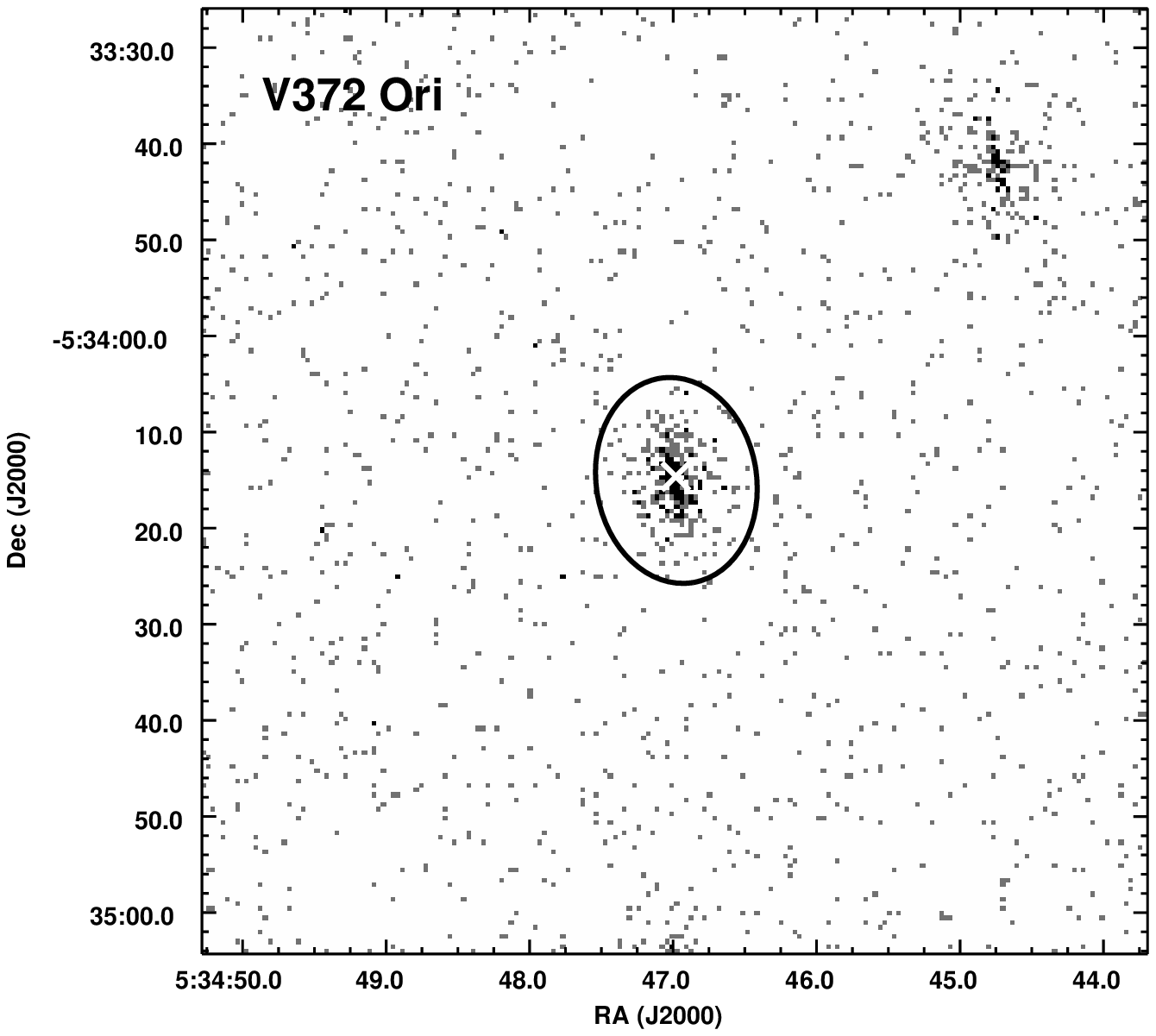}}
      }
    \mbox{
      {\includegraphics[angle=0,scale=0.43]{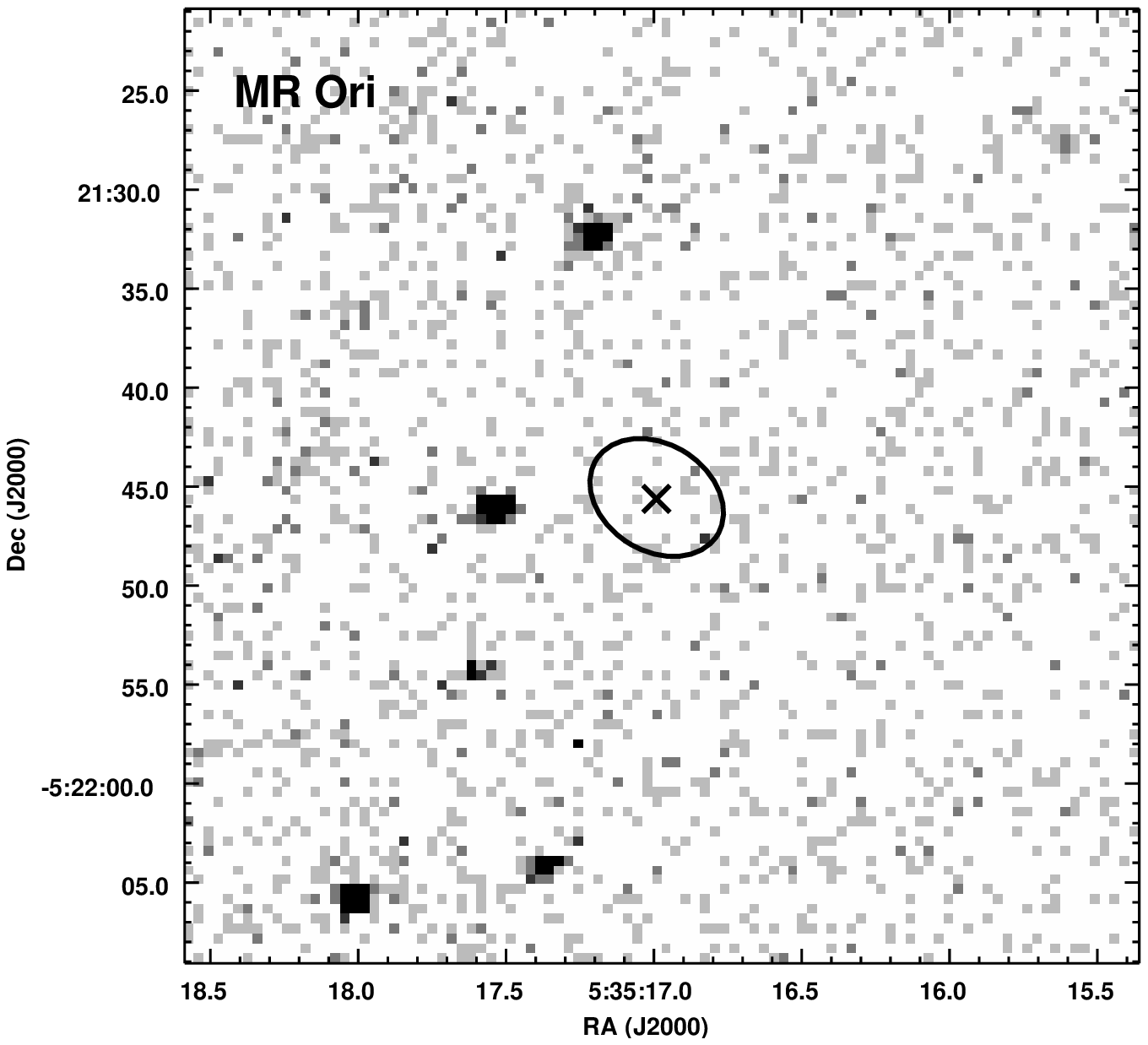}} \quad
      {\includegraphics[angle=0,scale=0.43]{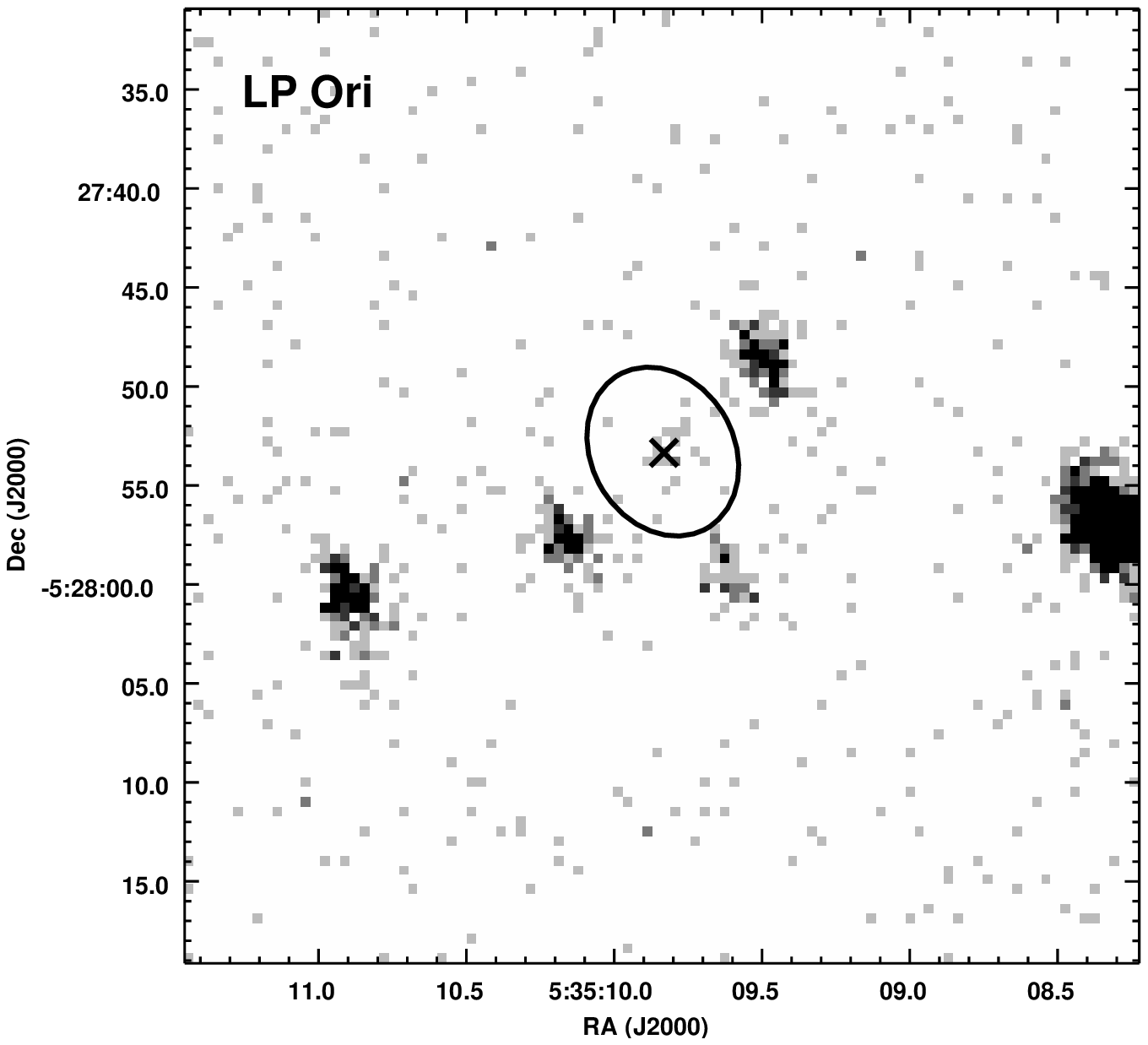}} 
      }
    \mbox{
      {\includegraphics[angle=0,scale=0.43]{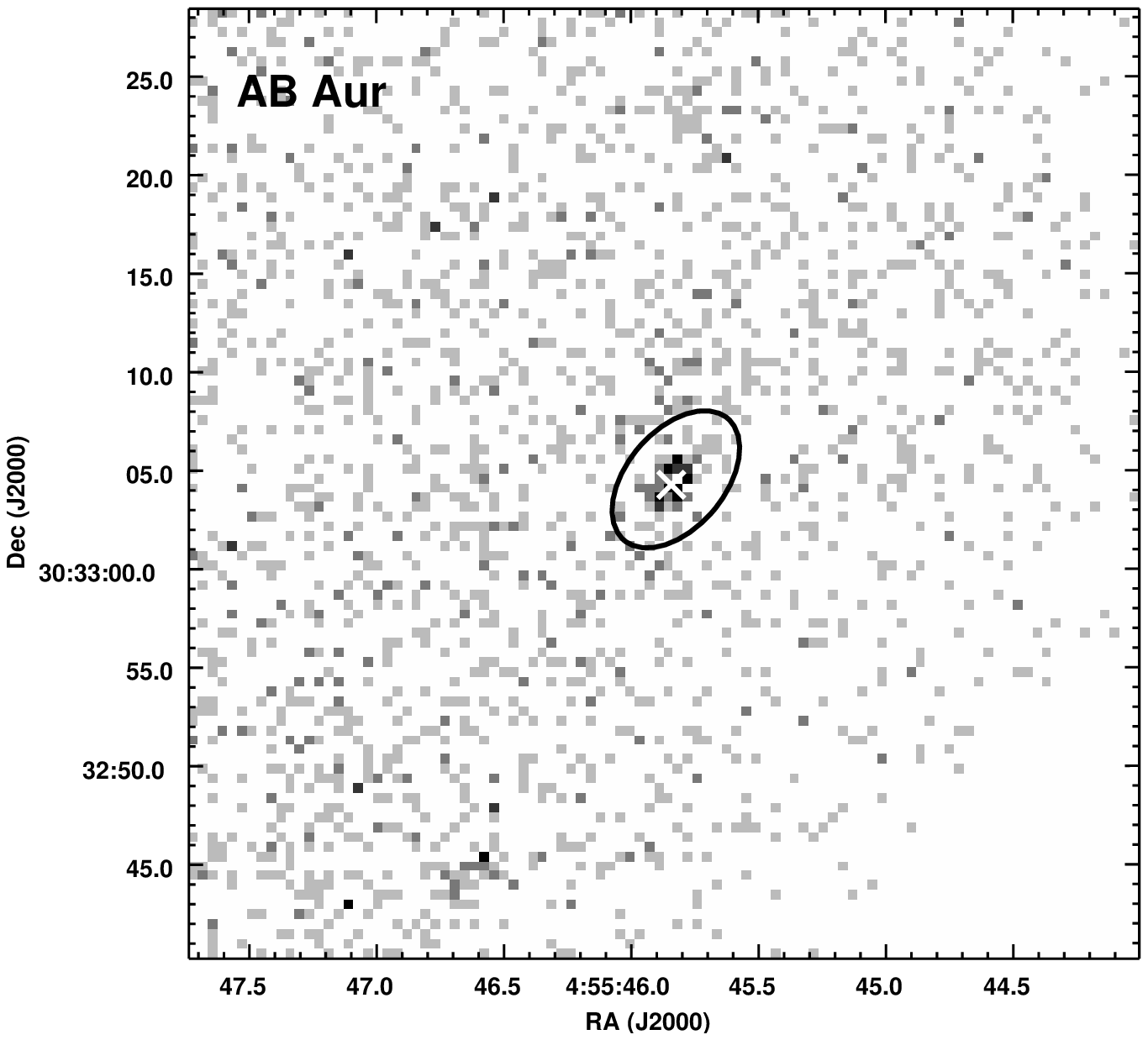}} \quad
       }
    \caption{Maps of the five new observed Herbig sources with Chandra (\S
      \ref{detect}). V361 Ori, V372 Ori, and AB Aur are detected and MR Ori
      and LP Ori are not detected (Table 3). The ellipses mark the PSF and the crosses the optical positions of the sources.\label{maps}}.
  \end{center}
\end{figure}

\clearpage

\begin{figure}
\includegraphics[angle=0]{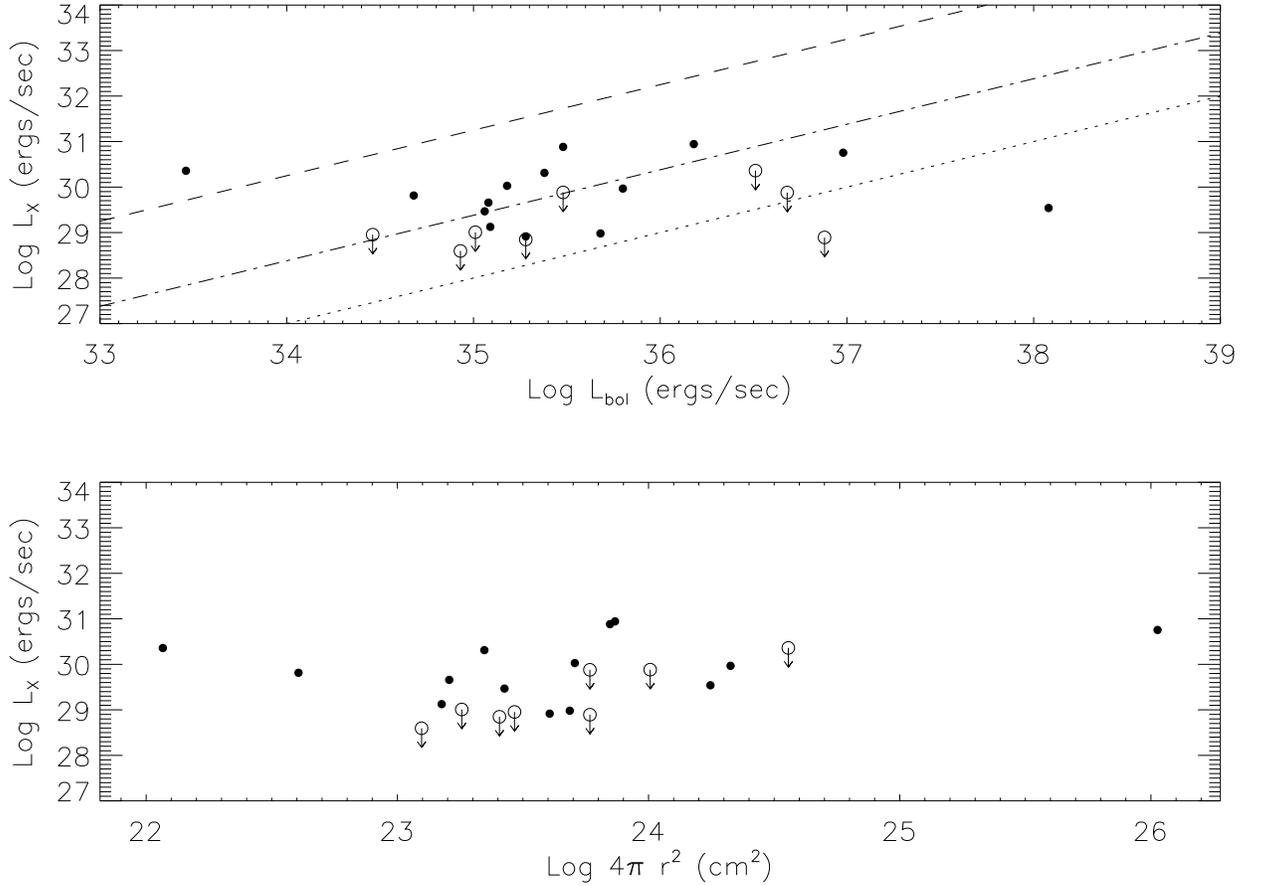}
\caption{\textit{Top:} X-Ray luminosity Log $L_X$ of detected and non-detected
  sources versus Log $L_{bol}$. For non-detected sources we give the upper
  limits, marked with the down pointing arrows. The
  straight lines correspond to Log ($L_X/L_{bol}$) = -3.75 (dashed), -5.62
  (dash-dot), and -7.0 (dots). \textit{Bottom:} Log $L_X$ versus the stellar
  surface area. 
\label{lbol}}   
\end{figure}

\clearpage

\begin{figure}
\includegraphics[angle=0]{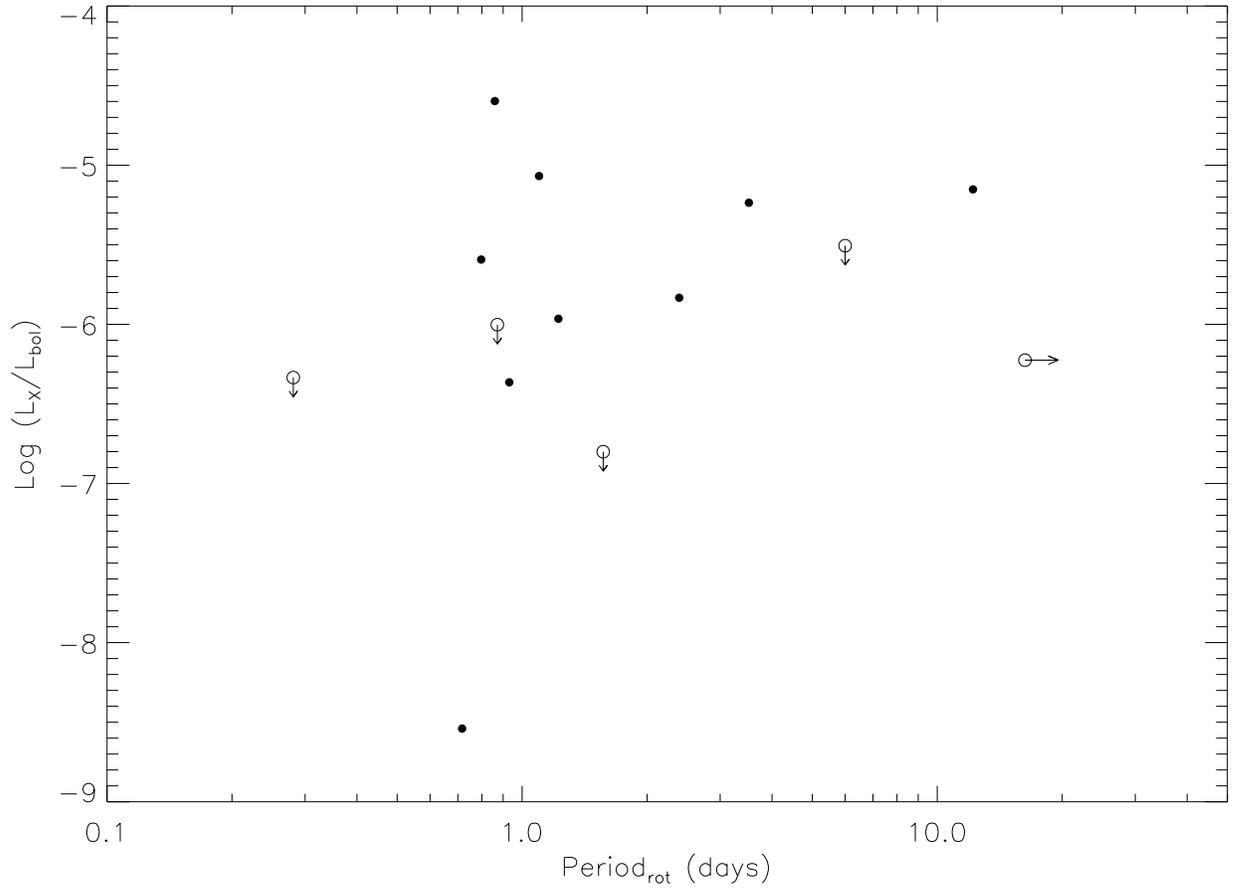}
\caption{The luminosity ratio Log $L_X/L_{bol}$ versus the rotational period
  of the stars with known \textit{v sin i}. The down pointing arrows mark the upper
  limits of the luminosity ratio. The right pointing arrow marks the lower
  limit of the rotational period.
\label{correl}}   
\end{figure}

\clearpage

\begin{figure}
\includegraphics[angle=-90,scale=0.65]{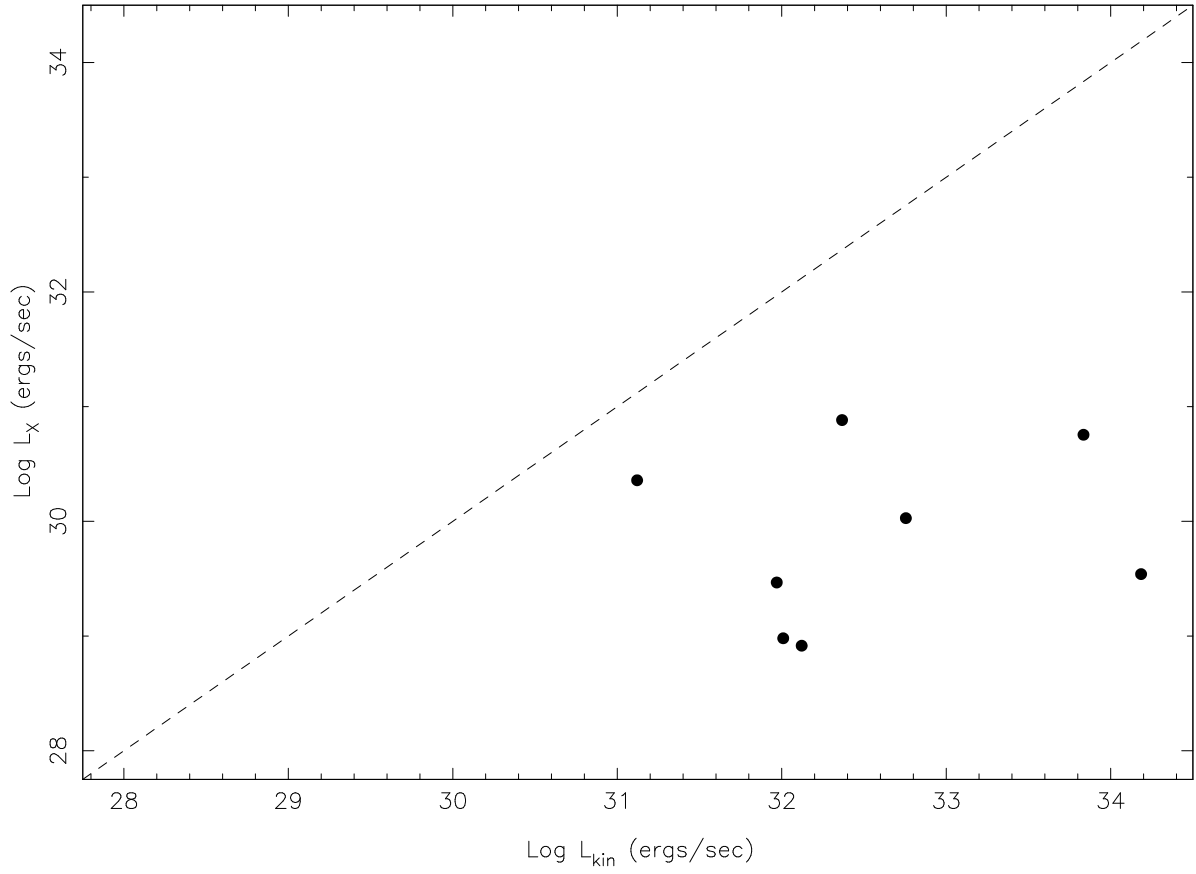}
\caption{Log $L_X$ versus wind Log $L_{kin}$; the straight line corresponds to
  $L_X$ = $L_{kin}$. 
\label{correl2}}   
\end{figure}

\clearpage

\begin{figure}
\centering
\includegraphics[angle=-90,scale=0.65]{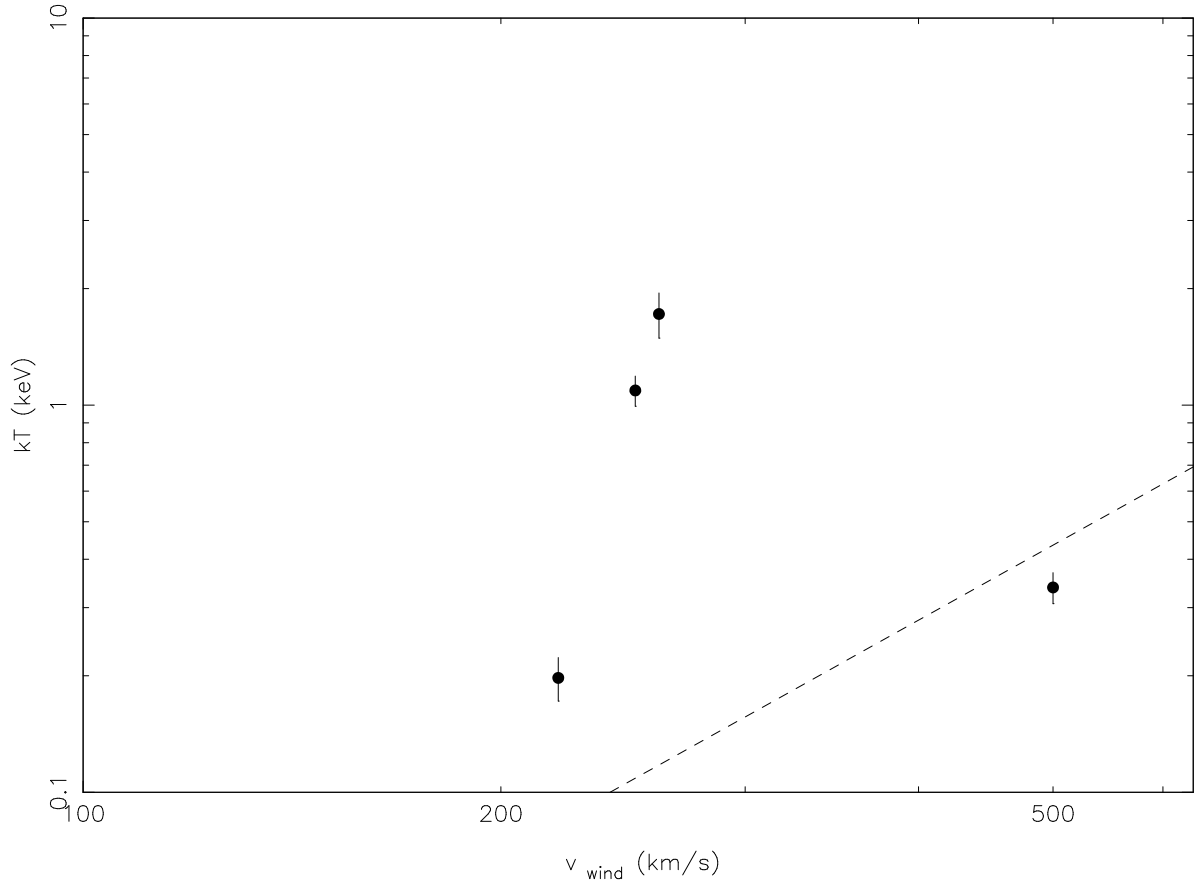}
\caption{Plasma temperature kT versus wind velocity \textit{v wind}; the straight line is deduced from the wind kinetic energy kT =1/3($\frac{1}{2} m_p v_{wind}^2$), where m$_p$ is the proton mass. HD 104237 is the source below the dashed line.  
\label{kt}}
\end{figure}

\clearpage

\begin{figure}
\centering
\includegraphics[angle=0,width=6.in]{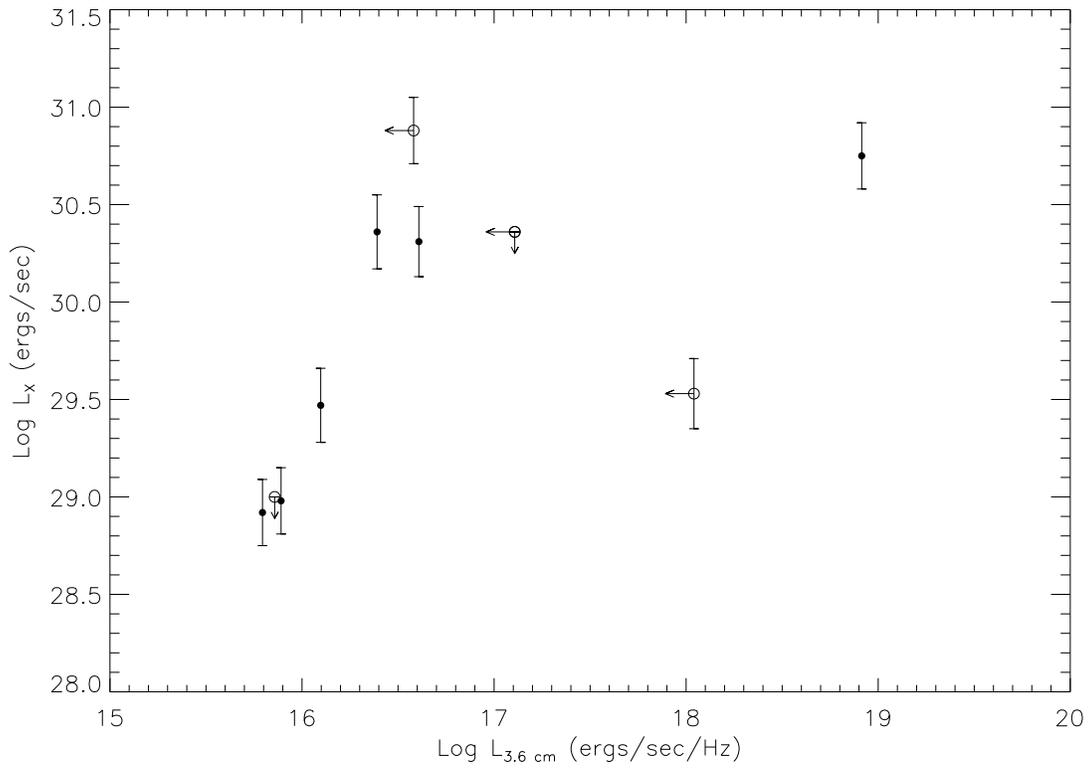}
\caption{Variation of Log $L_X$ vs. radio Log $L_{3.6 cm}$. The down pointing arrows mark the upper
  limits $L_X$. The left pointing arrows mark the upper
  limit of $L_{3.6 cm}$. \label{radio}}
\end{figure}

\clearpage
\begin{figure}
\includegraphics[angle=-90,scale=0.65]{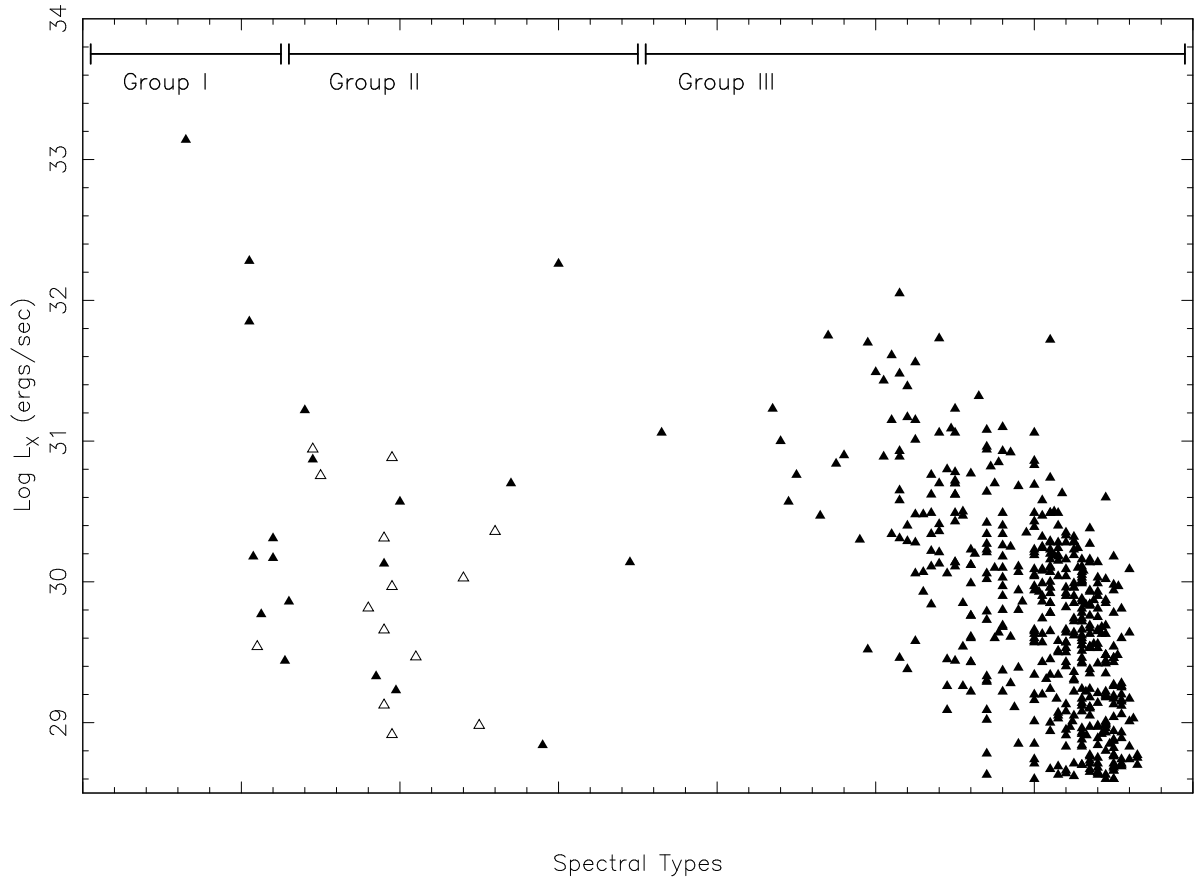}
\caption{X-Ray luminosity from Chandra observations
  variation with the spectral type: \textit{triangles} only the detected HAEBE data (Table 3) and \textit{filled-triangles} COUP observations. \label{spt}} 
\end{figure}

\clearpage
\begin{figure}
\includegraphics[angle=90,scale=0.65]{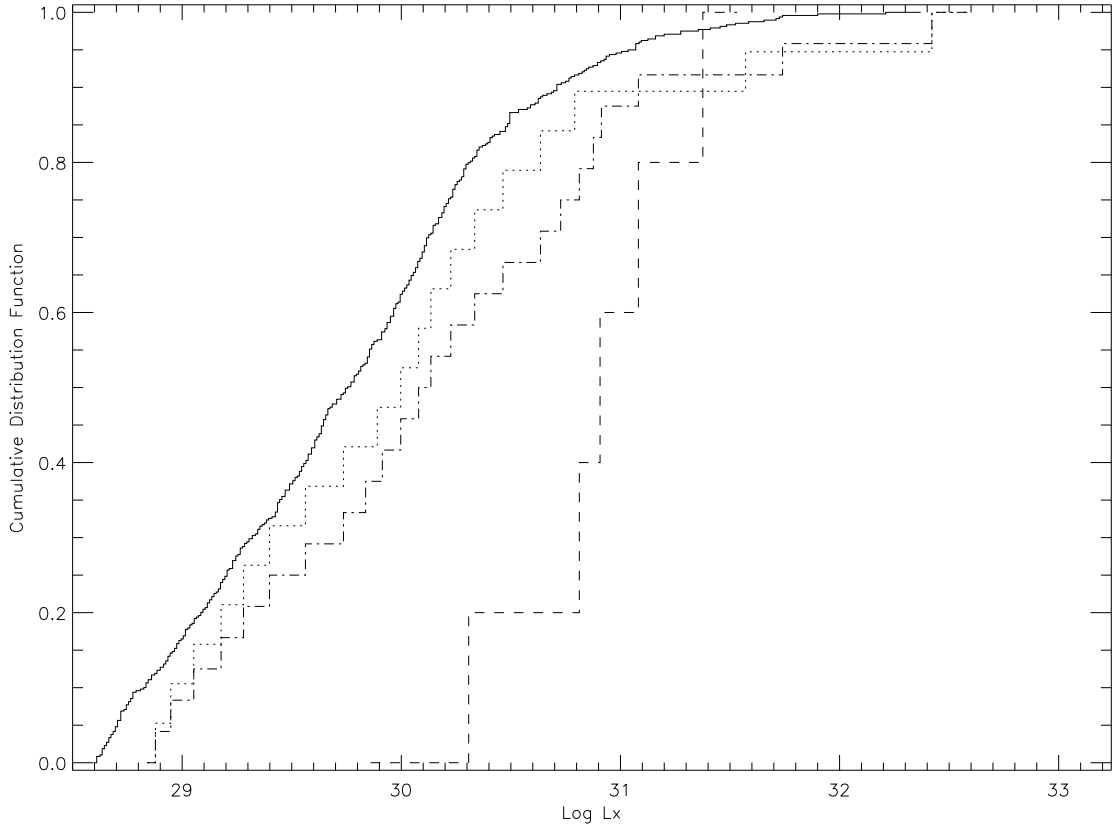}
\caption{$L_X$ Cumulative distribution function from Figure 7 of different
  samples: TTS (solid line), HAEBE (dash-dot line), Herbig Ae stars (dot line), and
  Herbig Be stars (dashed line).\label{cdf}}
\end{figure}



\clearpage
\begin{deluxetable}{lccccccccc}
\tabletypesize{\scriptsize}
\tablecaption{The physical parameters of the 22 Herbig AeBe stars.\label{tbl1}}
\tablewidth{0pt}
\tablehead{
\colhead{Object} & \colhead{Spectral\tablenotemark{a}} &
\colhead{Distance\tablenotemark{a}} & \colhead{Log
  $L_{bol}$\tablenotemark{a,b}} & \colhead{$S_{3.6 cm}$\tablenotemark{d}} & \colhead{Log T$_{eff}$}\tablenotemark{a,b} & \colhead{$v_{rot}$ \textit{sin i}\tablenotemark{b}} &  \colhead{$v_{wind}$ \tablenotemark{a,b}} & \colhead{RA\tablenotemark{a,b,c}} & \colhead{DEC\tablenotemark{a,b,c}} \\
\colhead{} &  \colhead{Type} & \colhead{(pc)}  & \colhead{($ergs$ $s^{-1}$)} & \colhead{(mJy)} & \colhead{(K)} & \colhead{(km s$^{-1}$)} & \colhead{(km s$^{-1}$)} & \colhead{(J2000)} & \colhead{(J2000)} 
}
\startdata
MWC 297 & B1Ve & 250 & 38.08 & $<$8.78 & 4.52 & 380 & 350 & 18 27 39.6 & -03 49 52 \\
LP Ori & B2 & 460 & 36.68 & ... & 4.29 & 100 & ... & 05 35 09.83 & -05 27 53.33 \\ 
HD 147889 & B2 & 140 & 36.88 & ... & 4.34 & ... & ... & 16 25 24.31 & -24 27 56.56\\ 
V361 Ori & B4/5 & 460 & 36.18 & ... & 4.14 & 50 & ... & 05 35 31.43 & -05 25 16.40 \\
Z CMa & B5 & 1150 & 36.98 & 3.1 & 3.80 & $<$130 & 500 & 07 03 43.16 & -11 33 06.20 \\
Lkh$\alpha$ 25 & B7 & 800 & 36.51 & $<$0.10 & 4.05 & ... & 340 & 06 40 44.56 & +09 48 02.2 \\ 
BD+30 549 & B8V & 390 & 34.68 & ... & 4.08 & ... & ... & 03 29 19.77 & +31 24 57.04 \\
R CrA & A5II & 130 & 35.68 & 0.23 & 4.06 & ... & 150 & 19 01 53.65 & -36 57 07.62 \\
V380 Ori & B8/A1 & 460 & 35.48 & $<$0.09 & 3.97 & 200 & 260 & 05 36 25.43 & -06 42 57.70 \\
HD 97300 & B9V & 188 & 35.08 & ... & 4.03 & ... & ... & 11 09 50.01 & -76 36 47.72 \\
HD 100546 & B9V & 103$\pm$7 & 35.09 & ... & 4.04 & 65$\pm$5 & ... & 11 33 25.44 & -70 11 41.23 \\
HD 176386 & B9 & 130 & 35.28 & ... & 4.03 & ... & ... & 19 01 38.93 & -36 53 26.54 \\
HD 141569 & B9.5 & 99$^{+9}_{-8}$ & 34.93 & ... & 4.02 & 258$\pm$17 & ... & 15 49 57.74 & -03 55 16.36  \\ 
AB Aur & B9/A0V & 144$^{+23}_{-17}$ & 35.28 & 0.15 & 3.98 & 140 & 225 & 04 55 45.84 & +30 33 04.29 \\
V372 Ori & B9.5V & 460 & 35.8 & ... & 3.93 & 125 & ... & 05 34 46.98 & -05 34 14.60 \\
HD 150193 & A2IV & 150$^{+50}_{-30}$ & 35.01 & 0.16 & 4 & 100 & 130 & 16 40 17.92 & -23 53 45.18 \\
HD 163296 & A1 & 122$^{+17}_{-13}$ & 35.06 & 0.42 & 3.97 & 133$\pm$6 & 220 & 17 56 21.28 & -21 57 21.88 \\
MR Ori & A2V & 460 & 35.48 & ... & 3.93 & ... & ... &05 35 16.99 & -05 21 45.6 \\ 
TY CrA & B9 & 130 & 35.38 & 1.2 & 4.07 & 10 & ... & 19 01 40.83 & -36 52 33.88 \\
Elias 3-1 & A6 & 160 & 33.46 & 0.48 & 3.91 & ... & 250 & 04 18 40.60 & +28 19 16.7 \\
HD 104237 & A4 & 116$^{+8}_{-7}$ & 35.18 & ... & 3.93 & 12$\pm$2 & 500 & 12 00 05.08 & -78 11 34.56 \\
AK Sco & F5IV & 150$^{+40}_{-30}$ & 34.46 & ... & 3.81 & 18.5$\pm$1 & ... & 16 54 44.84 & -36 53 18.57
\enddata
\tablenotetext{a}{Ref. \citet{hillenbrand92}, \citet{the94},
  \citet{malfait98}, \citet{vda98}, \citet{natta00}, \citet{fuente02}, \citet{hamaguchi05}.}
\tablenotetext{b}{Ref. \citet{damiani94}, \citet{skinner93},
  \citet{mora01}.}
\tablenotetext{c}{Ref. This research has made use of the SIMBAD database,
 operated at CDS, Strasbourg, France.}
\tablenotetext{d}{Ref. $S_{3.6 cm}$ are from \citet{skinner93},
  \citet{forbrich06}, and \citet{natta04}.}
\end{deluxetable}

\clearpage
\begin{deluxetable}{ lccccc }
\tabletypesize{\scriptsize}
\tablecaption{Chandra observations.\label{tbl2}}
\tablewidth{0pt}
\tablehead{
\colhead{Object} & \colhead{Observation} & \colhead{Date} & \colhead{Detector} & \colhead{Exposure Time} & \colhead{Intended}
\\
\colhead{} & \colhead{ID} & \colhead{} & \colhead{} & \colhead{(s)} & \colhead{Target}
}
\startdata
MWC 297 & 1883 & 2001-09-21 &  ACIS-I & 7731.518 & Yes \\
LP Ori & 3498 &  2003-01-21 &  ACIS-I & 66811.9 & No\\ 
HD 147889 & 618 & 2000-06-22 & ACIS-I & 3000 & No \\
V361 Ori & 3498 & 2003-01-21 & ACIS-I & 66811.9 & No\\
Z CMa & 3751 & 2003-12-07 &   ACIS-S & 37872.4 & Yes \\
Lkh$\alpha$ 25 & 2550  &   2002-02-09  &  ACIS-I  & 48755.2 & No \\ 
BD+30 549 & 642 & 2000-07-12 & ACIS-I & 43907.2 & No\\
R CrA & 3499   &  2003-06-26 &   ACIS-I & 37976.736 & Yes \\
V 380 Ori & 21 & 2000-10-08 & ACIS-S & 19610.0 & No \\
HD 97300 & 1867 & 2001-07-02 & ACIS-I & 66992.318 & No \\
HD 100546 & 3427 & 2002-02-04  &  ACIS-I & 2662.4 & Yes \\
HD 176386 & 3499 & 2003-06-26  &  ACIS-I  & 37976.736 & No \\
HD 141569 & 981 & 2001-06-23   & ACIS-I & 2803.60 & Yes \\ 
AB Aur & 3755  & 2003-11-27 &   ACIS-S & 99622.8 & No \\
V372 Ori & 2548 & 2002-09-06 & ACIS-I & 47414.0 & No \\
HD 150193 & 982  & 2001-08-19 & ACIS-I & 2806.8 & Yes \\
HD 163296 &  3733 & 2003-08-10 & ACIS-S & 19988.4 & Yes \\
MR Ori & 2568  & 2002-02-19 & ACIS-S  & 46928.0 & No \\ 
TY CrA & 3499 & 2003-06-26 & ACIS-I & 37976.736 & No \\
Elias 3-1 & 3364  & 2002-03-07 &   ACIS-S  & 17811.6 & No \\
HD 104237 & 3428  & 2002-02-04  &  ACIS-I & 2714.0 & Yes \\
AK Sco & 983 &  2001-08-19 &   ACIS-I & 3002.0 & Yes 
\enddata
\end{deluxetable}

\clearpage

\begin{deluxetable}{ lccccccc }
\tabletypesize{\scriptsize}
\tablecaption{X-Ray Detection of Herbig AeBe stars.\label{tbl3}}
\tablewidth{0pt}
\tablehead{
\colhead{Object\tablenotemark{a}} & \colhead{S/N}  & \colhead{Count Rate} & PSF & \colhead{Log $L_{X}$} & \colhead{Intended} & \colhead{Log L$_X$/L$_{bol}$} & \colhead{Detected}
\\
\colhead{} & \colhead{} & \colhead{(cts/ks)} &   & \colhead{($ergs$ $s^{-1}$)} & \colhead{Target} & & 
}
\startdata
MWC 297 & 4.54 & 2.88$\pm$0.52 & 0.9 & 29.53$\pm$0.18 & Yes & -8.54 & Yes \\
\underline{LP Ori}    & 0.78 & $<$ 0.027 & 0.9 & $<$ 29.87 & No & $<$ -6.8 & No\\
HD 147889 & 1.28 & $<$ 0.62 & 0.9 & $<$ 28.89 & No & $<$ -7.98 & No \\
\underline{V361 Ori}  & 54.64 & 44.59$\pm$0.56 & 0.9 & 30.94$\pm$0.17 & No & -5.24 & Yes\\
Z CMa     & 6.33 & 1.08$\pm$0.15 & 0.9 & 30.75$\pm$0.17 & Yes & -6.14 & Yes \\
Lkh$\alpha$ 25 & 0.98 & $<$ 0.04  & 0.9 & $<$ 30.36 &  No  & $<$ -6.15 & No \\ 
BD+30 549 & 4.92 & 1.05$\pm$0.13 & 0.6 & 29.81$\pm$0.16 & No & -4.87 & Yes \\
R CrA     & 10.73 & 3.04$\pm$0.18 & 0.9 & 28.98$\pm$0.17 & Yes & -6.69 & Yes \\
V380 Ori & 30.49 & 47.36$\pm$1.08 & 0.9 & 30.88$\pm$0.17 & No & -4.60 & Yes \\
HD 97300  & 32.76 & 16.03$\pm$0.34 & 0.9 & 29.66$\pm$0.18 & No & -5.4 & Yes \\
HD 100546 & 5.90 & 13.11$\pm$1.96 & 0.9 & 29.13$\pm$0.18 & Yes & -6.1 & Yes \\
HD 176386 & 1.14 & $<$0.05 & 0.9 & $<$ 28.84  & No & $<$ -6.43 & No \\
HD 141569 & 1.26 & $<$ 0.66 & 0.9 & $<$ 28.59 & Yes & $<$ -6.33 & No \\
\underline{AB Aur}    & 5.28 & 0.59$\pm$0.13 & 0.55 & 28.92$\pm$0.17 & No & -6.36 & Yes \\
\underline{V372 Ori}  & 17.55 & 6.43$\pm$0.28 & 0.9 & 29.97$\pm$0.17 & No & -5.83 & Yes \\
HD 150193 & 1.95 & $<$1.4 & 0.5 & $<$ 29.00 & Yes & $<$ -6.00 & No \\
HD 163296 & 34.28 & 53.70$\pm$1.18 & 0.9 & 29.47$\pm$0.19 & Yes & -5.59 & Yes \\
\underline{MR Ori} & 0.62 & $<$ 0.07 & 0.9  & $<$ 29.88 & No  & $<$ -5.59 & No \\ 
TY CrA    & 71.55 & 134.88$\pm$1.31 & 0.9 & 30.31$\pm$0.18 & No & -5.07 & Yes \\
Elias 3-1 & 39.87 & 89.26$\pm$1.40 & 0.9 & 30.36$\pm$0.19 & No & -3.10 & Yes \\
HD 104237 & 20.48 & 154.34$\pm$5.67 & 0.9 & 30.03$\pm$0.18 & Yes & -5.15 & Yes \\
AK Sco    & 1.31 & $<$ 0.63 & 0.9 & $<$ 28.95 & Yes & $<$ -5.50 & No  
\enddata
\tablenotetext{a}{The underlined sources' observations are reported in this
  study for
  the first time (\S \ref{detect}).}
\end{deluxetable}

\clearpage

\begin{deluxetable}{lcccc}
\tabletypesize{\scriptsize}
\tablecaption{X-ray parameters from the spectral analysis.\label{tbl4}}
\tablewidth{0pt}
\tablehead{
\colhead{Object} & \colhead{Model} & \colhead{T\tablenotemark{a}} & \colhead{N$_{H_2}$(10$^{22}$ cm$^{-2}$)} & \colhead{$\chi^2_r$ (dof)\tablenotemark{b}}
\\
\colhead{} &  \colhead{} & \colhead{(kev)} & \colhead{} & \colhead{} 
}
\startdata
V361 Ori & 2T & 0.95,3.08 & 0.13$\pm$0.03 & 0.38(273) \\
V380 Ori & 2T & 1.12,2.31 & 0.15$\pm$0.07 & 0.44(121) \\
HD 97300 & 2T & 0.86,2.91 & 0.31$\pm$0.15 & 0.17(310)\\
V372 Ori & 1T & 0.97 & 0.18$\pm$0.01 & 1.03(23) \\ 
HD 163296 & 1T & 0.39 & 0.06$\pm$0.05 & 0.19(542) \\
TY CrA & 2T & 0.79,2.07 & 0.21$\pm$0.04 & 0.33(303) \\
Elias 3-1 & 1T & 2.18 & 0.67$\pm$0.18 & 0.21(360)\\
HD 104237 & 1T & 0.67 & 0.38$\pm$0.07 & 0.13(430)
\enddata
\tablenotetext{a}{Fitted temperatures using one- or two-temperature MEKAL
  models, based on the best fit model or $\chi^2$.}
\tablenotetext{b}{Reduced $\chi^2$ level and the number of degrees of freedom in parentheses.}
\end{deluxetable}

\clearpage

\begin{deluxetable}{cccc}
\tabletypesize{\scriptsize}
\tablecaption{K-S and WRS tests results of two samples being
  drawn from a same parent distribution.}
\tablewidth{0pt}
\tablehead{
\colhead{Test} & \colhead{Datasets\tablenotemark{a}} & \colhead{K-S test probability\tablenotemark{b} ($\%$)} & \colhead{WRS test probability\tablenotemark{b} ($\%$)}
\\
}
\startdata
1\tablenotemark{c} & sub-Group III -- sub-Group III & 99.99 & 95 \\
2 & Group I -- Group II & 55 & 28 \\ 
3 & Group II -- Group III & 20 & 6 \\ 
4 & Group I -- Group III & 9 & 2 \\ 
5 & Herbig Ae -- Group III & 88 & 42 \\ 
\enddata
\tablenotetext{a}{See \S \ref{ONC}: Group I ($<$B3), Group II (B3-F5), Group
  III ($>$ F5)}. 
\tablenotetext{b}{Small values of the probability show that the distributions of the 2 datasets are significantly different.}
\tablenotetext{c}{Splitting randomly Group III in half to make 2
  sub-Groups; taking then the tests on these 2 sub-Groups allows to check the robustness of our tests.}
\end{deluxetable}


\end{document}